\documentclass[preprint2,numberedappendix]{emulateapj-rtx4}

\usepackage{graphicx}
\usepackage{natbib}

\usepackage{bm,times,color}
\graphicspath{{./fig/}{./png/}}
\usepackage{amsmath}

\newcommand{\EQ}{\begin{equation}}
\newcommand{\EN}{\end{equation}}
\newcommand{\EQA}{\begin{eqnarray}}
\newcommand{\ENA}{\end{eqnarray}}

\newcommand{\Eq}[1]{Equation~(\ref{#1})}

\newcommand{\Eqss}[2]{Equations~(\ref{#1})--(\ref{#2})}

\newcommand{\Sec}[1]{Section~\ref{#1}}

\newcommand{\Fig}[1]{Figure~\ref{#1}}

\newcommand{\Tab}[1]{Table~\ref{#1}}

\newcommand{\bra}[1]{\langle #1\rangle}

{}
{}
{}

{}
{}
{}
{}
{}
{}
{}
{}
{}
{}
{}
{}
{}
{}
{}
{}
{}
{}
{}
{}
{}

{}
{}
{}

{}

{}
{}

\newcommand{\eee}{\hat{\mbox{\boldmath $e$}} {}}

\newcommand{\yy}{\mbox{\boldmath $y$} {}}

\newcommand{\kk}{\bm{k}}

\newcommand{\xx}{\bm{x}}

\newcommand{\bb}{\bm{b}}
\newcommand{\BB}{\bm{B}}

\newcommand{\uu}{\mbox{\boldmath $u$} {}}

\newcommand{\JJ}{\mbox{\boldmath $J$} {}}

\newcommand{\AAA}{\mbox{\boldmath $A$} {}}

\newcommand{\ff}{\mbox{\boldmath $f$} {}}

\newcommand{\nab}{\mbox{\boldmath $\nabla$} {}}
\newcommand{\OO}{\bm{\Omega}}

\newcommand{\RRRR}{\mbox{\boldmath ${\sf R}$} {}}
\newcommand{\SSSS}{\mbox{\boldmath ${\sf S}$} {}}

\newcommand{\ii}{{\rm i}}

\newcommand{\DD}{{\rm D} {}}

\newcommand{\dd}{{\rm d} {}}

\newcommand{\const}{{\rm const}  {}}

\def\la{\mathrel{\mathchoice {\vcenter{\offinterlineskip\halign{\hfil
$\displaystyle##$\hfil\cr<\cr\sim\cr}}}
{\vcenter{\offinterlineskip\halign{\hfil$\textstyle##$\hfil\cr<\cr\sim\cr}}}
{\vcenter{\offinterlineskip\halign{\hfil$\scriptstyle##$\hfil\cr<\cr\sim\cr}}}
{\vcenter{\offinterlineskip\halign{\hfil$\scriptscriptstyle##$\hfil\cr<\cr\sim\cr}}}}}

\def\Co{\mbox{\rm Co}}

\def\Sc{\mbox{\rm Sc}}

\def\Pm{\mbox{\rm Pr}_M}
\def\Rm{\mbox{\rm Re}_M}

\def\Rey{\mbox{\rm Re}}

\def\Co{\mbox{\rm Co}}

\def\EK{E_{\rm K}}
\def\EM{E_{\rm M}}

\def\cs{c_{\rm s}}

\def\kf{k_{\rm f}}

\def\EM{E_{\rm M}}

\def\epsK{\epsilon_{\it K}}
\def\epsM{\epsilon_{\it M}}
\def\epsC{\epsilon_{\it C}}
\def\epsT{\epsilon_{\it T}}

\def\urms{u_{\rm rms}}

\def\brms{b_{\rm rms}}

\def\half{{\textstyle{1\over2}}}

\def\onethird{{\textstyle{1\over3}}}

\newcommand{\yjgr}[3]{ #1, {J.\ Geophys.\ Res.,} {#2}, #3}

\newcommand{\ynjp}[3]{ #1, {New J. Phys.,} {#2}, #3}
\newcommand{\yapj}[3]{ #1, {ApJ,} {#2}, #3}

\newcommand{\yapjl}[3]{ #1, {ApJ,} {#2}, #3}

\newcommand{\yan}[3]{ #1, {Astron.\ Nachr.,} {#2}, #3}

\newcommand{\yana}[3]{ #1, {A\&A,} {#2}, #3}

\newcommand{\yjfm}[3]{ #1, {J.\ Fluid Mech.,} {#2}, #3}

\newcommand{\ypf}[3]{ #1, {Phys.\ Fluids,} {#2}, #3}

\newcommand{\ypp}[3]{ #1, {Phys.\ Plasmas,} {#2}, #3}

\newcommand{\yjetp}[3]{ #1, {Sov.\ Phys.\ JETP,} {#2}, #3}

\newcommand{\yprl}[3]{ #1, {Phys.\ Rev.\ Lett.,} {#2}, #3}

\newcommand{\ymn}[3]{ #1, {MNRAS,} {#2}, #3}
\newcommand{\ynat}[3]{ #1, {Nature,} {#2}, #3}

\newcommand{\yprd}[3]{ #1, {Phys.\ Rev.\ D,} {#2}, #3}
\newcommand{\ypre}[3]{ #1, {Phys.\ Rev.\ E,} {#2}, #3}

\newcommand{\yjour}[4]{ #1, {#2}, {#3}, #4}

\newcommand{\yproc}[5]{ #1, in {#3}, ed.\ #4 (#5), #2}

\begin{document}
\preprint{NORDITA-2014-52}

\title{Magnetic Prandtl number dependence of kinetic-to-magnetic dissipation ratio}

\author{Axel Brandenburg}
\affil{
Nordita, KTH Royal Institute of Technology and Stockholm University,
SE-10691 Stockholm, Sweden\\
Department of Astronomy, Stockholm University, SE-10691 Stockholm, Sweden}

\date{\today,~ $ $Revision: 1.62 $ $}

\begin{abstract}
Using direct numerical simulations of three-dimensional hydromagnetic
turbulence, either with helical or non-helical forcing, we show that
the ratio of kinetic-to-magnetic energy dissipation always increases
with the magnetic Prandtl number, i.e., the ratio of kinematic
viscosity to magnetic diffusivity.
This dependence can always be approximated by a power law, but the
exponent is not the same in all cases.
For non-helical turbulence, the exponent is around 1/3,
while for helical turbulence it is between 0.6 and 2/3.
In the statistically steady state, the rate of the energy conversion from
kinetic into magnetic by the dynamo must be equal to the Joule dissipation rate.
We emphasize that for both small-scale and large-scale dynamos, the efficiency
of energy conversion depends sensitively on the magnetic Prandtl number,
and thus on the microphysical dissipation process.
To understand this behavior, we also study shell models of turbulence
and one-dimensional passive and active scalar models.
We conclude that the magnetic Prandtl number dependence is qualitatively
best reproduced in the one-dimensional model as a result of dissipation
via localized Alfv\'en kinks.
\end{abstract}


\section{Introduction}

One of the central paradigms of hydrodynamic turbulence is the equivalence of
large-scale energy injection and small-scale dissipation into heat
through viscosity---regardless of how small its value.
This is believed also to apply under conditions of astrophysically large
Reynolds numbers, when the
microphysical viscosity becomes very small compared with the product of
the physical scales and velocities of the system.
Dramatic examples are quasars, whose luminosities are equal to that of
a hundred galaxies and this emission is caused just by the dissipation
of turbulence, even though the microphysical viscosity is extremely small.
The detailed physical processes are not well understood, but it is
now generally believed that they also involve magnetic fields
\citep{SS73,BH98}.

Indeed, magnetic fields provide an additional important pathway for
dissipating turbulent energy through Joule heating.
The heating rates for both viscous and Joule dissipation are
proportional to the microphysical values of viscosity $\nu$ and
magnetic diffusivity $\eta$, respectively.
The ratio of these coefficients is the magnetic Prandtl number, $\Pm=\nu/\eta$.
As these coefficients are decreasing, the velocity and magnetic field
gradients sharpen just enough so that the heating rates remain independent
of these coefficients.
For the magnetic case of Joule heating, the independence of the
magnetic Reynolds number was demonstrated by \cite{GN96} and \cite{Hendrix}
in connection with the coronal heating problem.
Over a range of magnetic Reynolds numbers, the approximate constancy
of Joule dissipation has also been seen in turbulent dynamo simulations
\citep{CHMB11}.

While this picture is appealing and seemingly well confirmed, at least
in special cases such as for fixed values of $\Pm$, questions have arisen
in those cases when the magnetic and fluid Reynolds numbers are changed in
such a way that their ratio changes.
Hydromagnetic turbulence simulations exhibiting dynamo action have shown
that the values of energy dissipation are then no longer constant, and
that their ratio scales with $\Pm$ \citep{Min07,B09,B11,B11an}.
Given that all of the energy that is eventually dissipated comes from the
forcing in the momentum equation, a change in the dissipation ratio
can only be a consequence of a change in the conversion of kinetic
to magnetic energy through the dynamo process.
Therefore, the dynamo process would be intimately linked to Joule
dissipation and one must therefore be concerned that it is also linked
to the physical or even numerical nature of energy dissipation.
This would be surprising, because dynamo action has frequently been
modeled in many astrophysical turbulence simulations by focusing on
the so-called ideal equations with numerical dissipation only where no
$\Pm$ can be defined.
Examples in the context of local accretion disk dynamo simulations can be
found in the papers by \cite{BNST95}, \cite{HGB96}, and \cite{SHGB96}.
This leads to an ignorance that is potentially dangerous if such simulations
are employed to make predictions concerning energy deposition in
accretion disks \citep[see discussion by][]{BKL97}.

There is some concern that the numerical results of \cite{B09,B11}
may not yet be in the asymptotic regime and that the $\Pm$ dependence might
disappear at sufficiently large values of $\Rey$.
However, two arguments against this possibility have now emerged.
First, there are analytic results in two-dimensional magnetohydrodynamics
(MHD) by \cite{TYB13} that demonstrate the boundedness of the mean-squared
current density and mean-squared vorticity in the limits of large and
small values of $\Pm$, respectively.
In fact, \cite{TYB13} also produce numerical scalings similar
to the results of \cite{B11,B11an}.
Second, MHD shell models of turbulence by \cite{PS10} for $\Pm>1$ show
a similar $\Pm$ dependence, which is remarkable because those models
can be extended to much larger values of $\Rm$ than what is currently
possible with DNS.

Thus, there is now mounting evidence for a genuine
dependence of the macroscopic properties of MHD turbulence on $\Pm$.
Another such dependence
has been discussed for some time in connection with non-helical turbulence
exhibiting small-scale dynamo action in the {\em kinematic} regime.
Note, however, that this no longer applies in the non-kinematic regime
\citep{B11}.
For a kinematic small-scale dynamo dynamo, the magnetic energy spectra
grow in an approximately shape-invariant fashion with an approximate
$k^{3/2}$ spectrum at small wavenumbers.
This spectrum was first predicted by \cite{Kaz68} in the case of a smooth flow.
This case corresponds to an idealized representation of turbulence at large
values of $\Pm$ \citep{Scheko02}, but this spectrum is apparently also
found at small values of $\Pm$ near unity \citep[see Figure~4 of][]{HBD04}.
Depending on the value of $\Pm$, the magnetic energy spectrum peaks at
wavenumbers either within the inertial range of the turbulence
or in the viscous subrange.
This has implications for the critical magnetic Reynolds number
for the onset of dynamo action \citep{RK97}.
As explained by \cite{BC04}, the velocity field is rough in the
inertial range.
This interpretation has been successfully applied when clarifying the reason
for an apparent divergence \citep{Scheko05} of the critical Reynolds number
above which dynamo action is possible \citep{Isk07,Scheko07}.

There has been a similar debate regarding the onset of the
magneto-rotational instability in local simulations of accretion disks
\citep{FP07,FPLH07}, where the instability was found not to be excited
for small values of $\Pm$.
However, these examples are restricted to the physics of small-scale
magnetic fields only.
If one allows large-scale fields to develop, e.g.,
by relaxing the restriction to closed or periodic boundary conditions,
this $\Pm$ dependence disappears \citep{KK11}.

In the following, we will be concerned with the fully
dynamic case where kinetic and magnetic energies are comparable.
The purpose of the present paper is to illuminate the problem of the
$\Pm$ dependence of the dissipation ratio through
a combination of different approaches to MHD turbulence
ranging from direct numerical simulations (DNS) of the MHD equations
in three dimensions and shell models of turbulence capturing aspects
of the spectral cascade, to a simple one-dimensional model of MHD
\citep[cf.][]{Tho68,Pou93,BNP14}.
This leads us to suggest that the $\Pm$ dependence found in
turbulent dynamo simulations is caused by the dominant influence of
dissipative structures on the turbulent cascade at larger scales.
These dissipative structures can be thought of as local
Alfv\'en kinks whose width is determined by the algebraic
mean of kinematic viscosity and magnetic diffusivity.

\section{Simulations of turbulent dynamos}
\label{TurbulentDynamos}

\subsection{Governing equations}

In this section, we consider forced MHD turbulence of a gas that can be
described by an isothermal equation of state, i.e., the gas pressure
$p$ is proportional to the gas density $\rho$ with $p=\rho\cs^2$, where
$\cs=\const$ is the isothermal sound speed.
We apply a forcing function $\ff$ that is either fully helical or non-helical.
In both cases, there is initially just a weak seed magnetic field,
which is then amplified by dynamo action.
In the former case with helicity, we obtain large-scale magnetic fields,
as were studied previously with similar setups \citep{B01,B09,Min07}, while
in the latter case only small-scale dynamo action is possible
\citep{CV00,HBD03,HBD04,Scheko04,B11}.
In some cases, we also include the Coriolis force to study
the effects of rotation.
We solve the governing equations in the form
\EQA
\label{DH}
{\DD\ln\rho\over\DD t}&=&-\nab\cdot\uu,\\
\label{Du}
{\DD\uu\over\DD t}&=&-\cs^2\nab\ln\rho-2\OO\times\uu+\ff
\nonumber \\
&&+\rho^{-1}\left[\JJ\times\BB+\nab\cdot(2\nu\rho\SSSS)\right],\\
\label{dA}
{\partial\AAA\over\partial t}&=&\uu\times\BB-\eta\mu_0\JJ,
\ENA
where $\DD/\DD t=\partial/\partial t+\uu\cdot\nab$ is the advective
derivative, $\uu$ is the velocity, $\BB=\nab\times\AAA$ is the magnetic
field, $\AAA$ is the magnetic vector potential, $\JJ=\nab\times\BB/\mu_0$
is the current density, $\mu_0$ is the vacuum permeability, and
\EQ
{\sf S}_{ij}=\half(u_{i,j}+u_{j,i})-\onethird\delta_{ij}\nab\cdot\uu
\EN
is the traceless rate-of-strain tensor.
It is useful to note that
\EQ
\rho^{-1}\nab\cdot(2\rho\SSSS)={\textstyle{4\over3}}\nab\nab\cdot\uu
-\nab\times\nab\times\uu+2\SSSS\cdot\nab\ln\rho,
\label{nab2S}
\EN
where we call attention to the presence of the $4/3$ factor which
will be relevant for irrotational flows.

We consider a triply periodic domain, so that the kinetic and magnetic
energy balance is described by
\EQ
{\dd\over\dd t}\bra{\rho\uu^2/2}=
\bra{p\nab\cdot\uu}+\bra{\uu\cdot(\JJ\times\BB)}
+\bra{\rho\uu\cdot\ff}-\bra{2\rho\nu\SSSS^2},
\EN
\EQ
{\dd\over\dd t}\bra{\BB^2/2\mu_0}=-\bra{\uu\cdot(\JJ\times\BB)}
-\bra{\eta\mu_0\JJ^2},
\EN
where $\SSSS^2={\sf S}_{ij}{\sf S}_{ji}$.
The total (kinetic plus magnetic) energy is sourced by
$\bra{\rho\uu\cdot\ff}$ and dissipated by the sum
of viscous and Joule dissipation, $\epsT=\epsK+\epsM$, with
\EQ
\epsK=\bra{2\rho\nu\SSSS^2}\quad\mbox{and}\quad
\epsM=\bra{\eta\mu_0\JJ^2}.
\EN
The terms $\bra{p\nab\cdot\uu}$ and $\bra{\uu\cdot(\JJ\times\BB)}$
characterize the work done by gas expansion and Lorentz force,
respectively.

\begin{figure}[t!]\begin{center}
\includegraphics[width=\columnwidth]{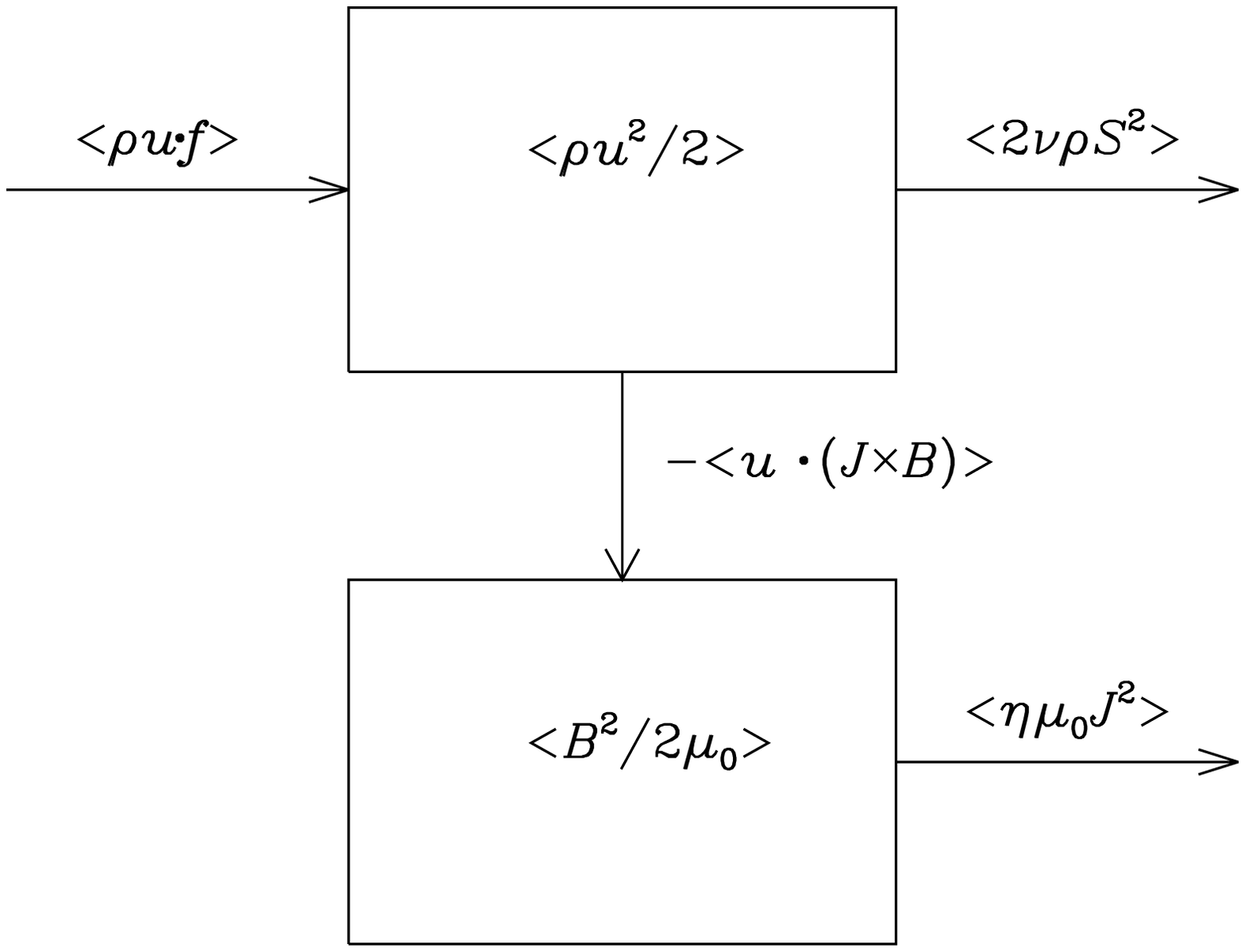}
\end{center}\caption[]{
Sketch showing the flow of energy injected by the forcing
$\bra{\rho\uu\cdot\ff}$ and eventually dissipated viscously and
resistively via the terms $\epsK$ and $\epsM$.
Note that in the steady state, $\epsM$ must be balanced by
$-\bra{\uu\cdot(\JJ\times\BB)}$.
}\label{psketch}\end{figure}

A sketch showing the transfers in and out of the two energy reservoirs,
$\EK=\bra{\rho\uu^2/2}$ and $\EM=\bra{\BB^2/2\mu_0}$,
is given in \Fig{psketch}.
From this it is clear that, in the steady state, the quantity
$-\bra{\uu\cdot(\JJ\times\BB)}$ must be positive 
and equal to $\bra{\eta\mu_0\JJ^2}$.

\subsection{The model}

We solve \Eqss{DH}{dA} with periodic boundary conditions using the
{\sc Pencil Code}\footnote{\url{http://pencil-code.googlecode.com/}},
which employs sixth order finite differences and a third order
accurate time stepping scheme.
For most of our runs, we choose a resolution of $512^3$ meshpoints.

In all cases, the amplitude of the forcing function is $f_0=0.02$,
which results in a Mach number $\urms/\cs$ of around 0.1.
Here, $\urms$ is the root-mean-square (rms) value of the
resulting velocity.
The simulations are further characterized by the fluid and magnetic
Reynolds numbers,
\EQ
\Rey=\urms/\nu\kf,\quad\Rm=\urms/\eta\kf,
\EN
so $\Pm=\Rm/\Rey$.
In cases with rotation, we also specify the Coriolis number,
\EQ
\Co=2\Omega/\urms\kf.
\EN
The energy supply for a helically driven dynamo is provided by the
forcing function $\ff = \ff(\xx,t)$, which is random in time
and defined as
\EQ
\ff(\xx,t)={\rm Re}\{N\ff_{\kk(t)}\exp[\ii\kk(t)\cdot\xx+\ii\phi(t)]\},
\label{ForcingFunction}
\EN
where $\xx$ is the position vector.
The wavevector $\kk(t)$ and the random phase
$-\pi<\phi(t)\le\pi$ change at every time step, so $\ff(\xx,t)$ is
$\delta$-correlated in time.
Therefore, the normalization factor $N$ has to be proportional to $\delta t^{-1/2}$,
where $\delta t$ is the length of the time step.
On dimensional grounds we choose
$N=f_0 c_{\rm s}(|\kk|c_{\rm s}/\delta t)^{1/2}$, where $f_0$ is a
non-dimensional forcing amplitude.
We use $f_0=0.02$, which results in a maximum Mach number of about 0.3
and an rms value of about 0.085.
At each timestep, we randomly select one of many possible wavevectors
in a certain range around a given forcing wavenumber with an
average value $k_{\rm f}$.
Transverse helical waves are produced via \citep{BS05b}
\begin{equation}
\ff_{\kk}=\RRRR\cdot\ff_{\kk}^{\rm(nohel)}\quad\mbox{with}\quad
{\sf R}_{ij}={\delta_{ij}-\ii\sigma\epsilon_{ijk}\hat{k}_k
\over\sqrt{1+\sigma^2}},
\label{eq: forcing}
\end{equation}
where $\sigma$ is a measure of the helicity of the forcing and
$\sigma=1$ for positive maximum helicity of the forcing function
and
\EQ
\ff_{\kk}^{\rm(nohel)}=
\left(\kk\times\eee\right)/\sqrt{\kk^2-(\kk\cdot\eee)^2}
\label{nohel_forcing}
\EN
is a non-helical forcing function, where $\eee$ is an arbitrary unit vector
that is not aligned with $\kk$; note that $|\ff_{\kk}|^2=1$ and
\EQ
\ff_{\kk}\cdot(\ii\kk\times\ff_{\kk})^*=2\sigma k/(1+\sigma^2),
\EN
so the relative helicity of the forcing function in real space is
$2\sigma/(1+\sigma^2)$; see \cite{CB13}.
In the cases mentioned below, we choose $\kf/k_1=3.1$ when $\sigma=1$,
so as to allow sufficient scale separation for the large-scale
field to develop, and $\kf/k_1=1.5$ when $\sigma=0$, where the issue
of scale separation is presumably less critical.

\begin{table*}\caption{
Summary of runs with $\Co=0$.
}\vspace{12pt}\centerline{\begin{tabular}{cccrrrccccccrrr}
\hline
\hline
Run & $\nu k_1/\cs$ & $\eta k_1/\cs$ & $\Rey~~$ & $\Rm$ & $\Pm$ & $\sigma$ &
$\urms/\cs$ & $\brms/\cs$ & $\epsK/\epsT$ & $\epsM/\epsT$ & $C_\epsilon$ &
$k_\nu/k_1$ & $k_\eta/k_1$ & res.~~ \\
\hline
%
A1  &$5.0\times10^{-4}$&$2.5\times10^{-5}$&  56&1123& 20.00&1& 0.087& 0.158&  0.81&  0.19&  1.83&  38& 247& $1024^3$ \\
A2  &$5.0\times10^{-4}$&$5.0\times10^{-5}$&  57& 568& 10.00&1& 0.088& 0.157&  0.76&  0.24&  1.80&  37& 156& $ 512^3$ \\
A3  &$5.0\times10^{-4}$&$1.0\times10^{-4}$&  57& 284&  5.00&1& 0.088& 0.157&  0.69&  0.31&  1.82&  36&  99& $ 512^3$ \\
A4  &$5.0\times10^{-5}$&$5.0\times10^{-5}$& 587& 587&  1.00&1& 0.091& 0.128&  0.39&  0.61&  1.75& 179& 201& $ 512^3$ \\
A5  &$5.0\times10^{-5}$&$2.5\times10^{-4}$& 606& 121&  0.20&1& 0.094& 0.155&  0.21&  0.79&  1.46& 150&  63& $ 512^3$ \\
A6  &$5.0\times10^{-5}$&$5.0\times10^{-4}$& 594&  59&  0.10&1& 0.092& 0.149&  0.15&  0.85&  1.60& 139&  38& $ 512^3$ \\
A7  &$5.0\times10^{-5}$&$1.0\times10^{-3}$& 581&  29&  0.05&1& 0.090& 0.149&  0.10&  0.90&  1.72& 125&  23& $ 512^3$ \\
\hline
%
B1  &$5.0\times10^{-5}$&$5.0\times10^{-5}$& 587& 587&  1.00&1& 0.091& 0.128&  0.39&  0.61&  1.75& 179& 201& $ 512^3$ \\
B2  &$2.5\times10^{-4}$&$5.0\times10^{-5}$& 117& 587&  5.00&1& 0.091& 0.159&  0.67&  0.33&  1.57&  60& 168& $ 512^3$ \\
B3  &$5.0\times10^{-4}$&$5.0\times10^{-5}$&  57& 568& 10.00&1& 0.088& 0.157&  0.76&  0.24&  1.80&  37& 156& $ 512^3$ \\
B4  &$1.0\times10^{-3}$&$5.0\times10^{-5}$&  27& 542& 20.00&1& 0.084& 0.155&  0.84&  0.16&  2.09&  23& 141& $ 512^3$ \\
\hline
%
C1  &$2.0\times10^{-5}$&$1.0\times10^{-4}$&1548& 310&  0.20&1& 0.096& 0.155&  0.19&  0.81&  1.30& 287& 124& $ 512^3$ \\
C2  &$2.0\times10^{-5}$&$2.0\times10^{-4}$&1532& 153&  0.10&1& 0.095& 0.149&  0.14&  0.87&  1.41& 268&  76& $ 512^3$ \\
C3  &$2.0\times10^{-5}$&$4.0\times10^{-4}$&1516&  76&  0.05&1& 0.094& 0.140&  0.10&  0.90&  1.47& 248&  46& $ 512^3$ \\
\hline
%
X1  &$5.0\times10^{-4}$&$5.0\times10^{-4}$&  56&  56&  1.00&0& 0.113& 0.043&  0.46&  0.54&  0.35&  28&  29& $ 256^3$ \\
X2  &$3.5\times10^{-5}$&$3.5\times10^{-4}$& 864&  86&  0.10&0& 0.121& 0.039&  0.18&  0.82&  0.26& 159&  41& $ 256^3$ \\
X3  &$7.0\times10^{-6}$&$3.5\times10^{-4}$&4179&  84&  0.02&0& 0.117& 0.041&  0.08&  0.92&  0.28& 422&  42& $ 512^3$ \\
\hline
%
Y1  &$1.0\times10^{-3}$&$5.0\times10^{-5}$&  55&1093& 20.00&0& 0.082& 0.070&  0.44&  0.56&  2.35&  16& 164& $ 512^3$ \\
Y2  &$5.0\times10^{-4}$&$5.0\times10^{-5}$& 121&1213& 10.00&0& 0.091& 0.066&  0.40&  0.60&  1.79&  27& 168& $ 512^3$ \\
Y3  &$2.5\times10^{-4}$&$5.0\times10^{-5}$& 245&1227&  5.00&0& 0.092& 0.066&  0.38&  0.62&  1.64&  44& 167& $ 512^3$ \\
Y4  &$1.0\times10^{-4}$&$5.0\times10^{-5}$& 647&1293&  2.00&0& 0.097& 0.065&  0.33&  0.67&  1.42&  85& 171& $ 512^3$ \\
Y5  &$5.0\times10^{-5}$&$5.0\times10^{-5}$&1293&1293&  1.00&0& 0.097& 0.062&  0.28&  0.72&  1.32& 135& 171& $ 512^3$ \\
Y6  &$2.5\times10^{-5}$&$5.0\times10^{-5}$&2533&1267&  0.50&0& 0.095& 0.063&  0.21&  0.79&  1.34& 210& 173& $ 512^3$ \\
Y7  &$1.0\times10^{-5}$&$5.0\times10^{-5}$&6400&1280&  0.20&0& 0.096& 0.059&  0.12&  0.88&  1.20& 356& 174& $ 512^3$ \\
\hline
\hline\label{Tsummary}\end{tabular}}
\end{table*}

\begin{figure}[t!]\begin{center}
\includegraphics[width=\columnwidth]{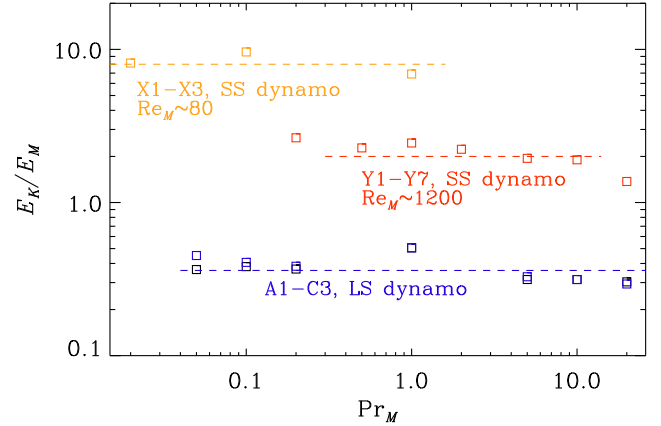}
\end{center}\caption[]{
Dependence of the ratio $\EK/\EM$ on $\Pm$ for
large-scale (LS) dynamos (solid blue line, Runs~A1--C3)
and small-scale (SS) dynamos (dashed orange and red lines, Runs~X1--Y7).
}\label{penergyKM}\end{figure}

\begin{figure}[t!]\begin{center}
\includegraphics[width=\columnwidth]{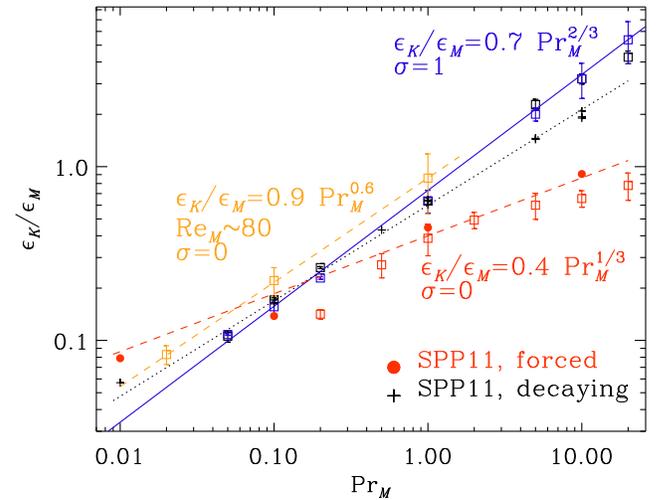}
\end{center}\caption[]{
Dependence of the dissipation ratio $\epsK/\epsM$ on $\Pm$ for
large-scale dynamos (solid blue line) and small-scale dynamos
(dashed orange and red lines).
The red filled symbols and black plus signs correspond to the results
of \cite{SPP11} for forced and decaying turbulence, respectively,
referred to as SPP11 in the legend.
}\label{pepsKM}\end{figure}

\subsection{Results}

In \Tab{Tsummary}, we present a summary of the runs discussed in
this paper.
As in \cite{B11}, $\epsK$ and $\epsM$ are normalized by their sum,
$\epsT=\epsK+\epsM$, which in turn is expressed in terms of the
non-dimensional quantity $C_\epsilon=a\epsT/\bra{\rho\urms^3\kf}$,
where $a=9\pi\sqrt{3}/4\approx12.2$ is a coefficient.
First of all, note that in all cases the energy ratio $\EK/\EM$
is roughly independent of $\Pm$ but it varies with $\Rm$, as was
demonstrated previously for the small-scale dynamo \citep{HBD03}.
For large-scale dynamos, the ratio $\EK/\EM$ is essentially equal to
$k_1/\kf$ \citep{B01}, which is around $0.3$ in the present case
(see \Fig{penergyKM}).
In \Fig{pepsKM}, we show the $\Pm$ dependence of $\epsK/\epsM$ for
$\sigma=1$ and 0.
The simulations show that for both $\sigma=1$ and 0, the ratio
$\epsK/\epsM$ scales with $\Pm$,
\EQ
\epsK/\epsM\propto\Pm^q,
\label{eps_scaling}
\EN
but the exponent is not always the same.
For $\sigma=1$, we find $q\approx2/3$ for both small and large values
of $\Pm$, while for $\sigma=0$, we find $q\approx0.6$ for $\Pm<1$
with $\Rey\approx80$ and $q\approx0.3$ for $\Pm>1$ with $\Rey\approx460$.
For large-scale dynamos ($\sigma=1$), a similar scaling was first found
for $\Pm\leq1$ \citep{Min07,B09}, and later also for $\Pm\geq1$ \citep{B11an}.
For $\Pm\leq1$, this scaling was also found for small-scale dynamos
\citep{B11}, but now we see that for $\Pm\geq1$ the slope is smaller.

Our results for $\Pm>1$ are compatible with those of \cite{SPP11},
who listed the kinetic and magnetic dissipation scales,
$\ell_K=(\nu^3/\epsK)^{1/4}$ and $\ell_M=(\eta^3/\epsM)^{1/4}$,
respectively, for their decaying and forced hydromagnetic simulations
at different values of $\Pm$.
Computing the dissipation ratio from their Table~1 as
$\epsK/\epsM=\Pm^3(\ell_K/\ell_M)^{-4}$, we find that their data for
non-helical decaying turbulence are well described by the formula
$\epsK/\epsM\approx0.6\,\Pm^{0.55}$.
For non-helically forced turbulence with $0.01\leq\Pm\leq10$, their
data agree perfectly with our fit $\epsK/\epsM\approx0.4\,\Pm^{1/3}$
(red filled symbols in \Fig{pepsKM}).
In their case, $\Rm$ increases with $\Pm$, but its value is generally
much larger than our values for $\Pm<1$.
This suggests that the $1/3$ scaling occurs for large enough magnetic
Reynolds numbers and that our steeper fit for $\Pm\leq1$ and the mismatch
at $\Pm=1$ is a consequence of small values of $\Rm$.

We emphasize that in view of \Fig{psketch}, the fraction of energy
that is being diverted to magnetic energy through dynamo action depends
on the term $-\bra{\uu\cdot(\JJ\times\BB)}$, and that this must be equal
to $\epsM$ in the statistically steady state.
This fraction is therefore $\epsM/\epsT$ and we may call it the
efficiency of the dynamo.
Remarkably, \Fig{pepsKM} shows that there is a $\Pm$ dependence of
the dynamo efficiency both with and without helicity.
The presence of helicity in the forcing function can lead to magnetic field
generation at the largest scale of the system.
It is therefore also referred to as a large-scale dynamo.
Non-helical forcing leads to magnetic fields on scales that are typically
somewhat smaller than the energy-carrying scale of the turbulent motions.

\begin{figure}[t!]\begin{center}
\includegraphics[width=\columnwidth]{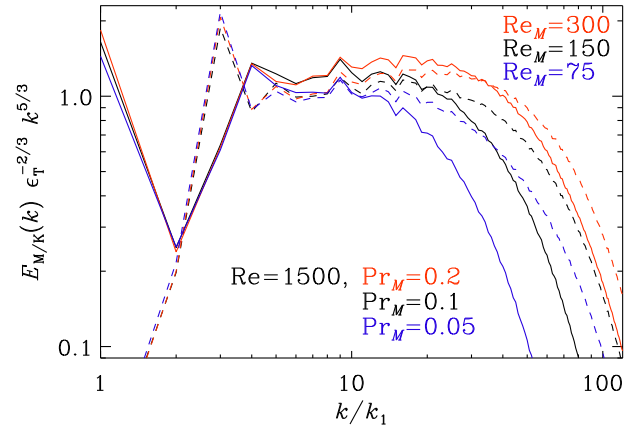}
\end{center}\caption[]{
Comparison of compensated kinetic (dashed) and magnetic (solid) energy
spectra for $\Pm=0.2$, 0.1, and 0.05 for helically forced turbulence.
}\label{pspec_comp_loPm}\end{figure}

One might be worried that these results are artifacts of the Reynolds
numbers still being too small and not yet in the asymptotic regime in which
a true $\Pm$-independence might be expected.
However, by comparing the energy spectra in at least some of the
cases indicates that there is indeed a short wavenumber range in which both
magnetic and kinetic spectra show an approximate $k^{-5/3}$ scaling with
wavenumber $k$ (see \Fig{pspec_comp_loPm}).
On the other hand, however, the presence of a residual slope may also
be regarded as evidence that none of the present simulations are yet in
the asymptotic regime.
Therefore, higher resolution simulations at larger Reynolds numbers
remain essential.
\\

The positive slope of the graph of $\epsK/\epsM$ versus $\nu/\eta$
indicates that a decrease of viscosity $\nu$ is not sufficiently
compensated by a sharpening of velocity gradients.
Likewise, a decrease of $\eta$ is not fully compensated by a corresponding
increase of $\JJ^2$.
In other words, as $\eta$ decreases, and thus $\Pm$ ($\gg1$) is further
increased, $\epsM$ still decreases and does not remain independent of $\eta$,
as would be the case for $\Pm=1$ \citep{Hendrix}.
This therefore leads to a residual increase of $\epsK/\epsM$.
This behavior was partially explained by the findings of \cite{B09,B11}
that for small values of $\Pm=\nu/\eta$, i.e., for
$\eta\gg\nu$, most of the spectral energy is dissipated through the magnetic
channel, leaving only a reduced amount of kinetic energy to be
dissipated, and therefore velocity gradients are not as sharp as in
the hydrodynamic case, $\epsK$ is reduced,
and $\epsK/\epsM$ decreases with decreasing values of $\Pm$.

Before closing the discussion on the $\Pm$ dependence in
three-dimensional MHD turbulence, let us comment on
the work term due to fluid expansion.
In all of the cases discussed here, $\bra{p\nab\cdot\uu}$ turns out to be
strongly fluctuating, although its time average is a very small fraction
of the total energy (0.02\%) for our low Mach number runs
(Mach numbers around 0.1).
There are indications, however, that $\bra{p\nab\cdot\uu}$ is negative
for $\Pm<1$ and positive for $\Pm>1$.

Given that there is currently no phenomenological explanation for the scaling
of $\epsK/\epsM$ given by \Eq{eps_scaling}, we must consider the possibility
that this scaling behavior is not generic and that different
scalings can be found in different situations.
To shed more light on the possible mechanisms that can explain these
scalings, we consider the results of an MHD shell model of turbulence.

\section{Shell models}

Shell models represent the dynamics of turbulence using scalar variables
for velocity and magnetic field along logarithmically spaced wavenumbers.
The governing equations resemble the original ones
with diffusion and forcing terms, as well as quadratic nonlinearities
that conserve the same invariants as the original equations: total
energy, cross helicity, and a proxy of magnetic helicity.
For a recent review of such models, see \cite{PSF13}.
The resulting set of equations can be written as
\EQA
\label{dundt}
{\partial u_n\over\partial t}&=&
\ii k_n\left[N_n(\uu,\uu)-N_n(\bb,\bb)\right]-\nu k_n^2u_n+F_n,\\
{\partial b_n\over\partial t}&=&
\ii k_n\left[M_n(\uu,\bb)-M_n(\bb,\uu)\right]-\eta k_n^2b_n,
\ENA
where $\uu=(u_1,u_2,...,u_N)$ and $\bb=(b_1,b_2,...,b_N)$ are
time-dependent complex vectors representing the state of the system
at wavenumbers $k_n=2^n$ with $n=1,2,...,N$.
The nonlinearities are given by \citep{BEO96,FS98}
\EQA
N_n(\xx,\yy)\!&=&\!x^*_{n+1}y^*_{n+2}
-\textstyle{1\over4}x^*_{n-1}y^*_{n+1}
-\textstyle{1\over8}x^*_{n-2}y^*_{n-1},\;\;
\\
M_n(\xx,\yy)\!&=&\!\textstyle{1\over6}(x^*_{n+1}y^*_{n+2}
-x^*_{n-1}y^*_{n+1}-x^*_{n-2}b^*_{n-1}).\;\;
\ENA
These equations preserve total energy, cross helicity, and
a proxy of magnetic helicity.
The only difference between \cite{BEO96} and \cite{FS98} is a
$12/5$ scaling factor in front of both nonlinear terms.
Models with these coefficients have been used to study the possibility of an
inverse cascade in the early universe \citep{BEO96,BEO97} and the onset properties
of small-scale dynamos \citep{FS98},
as well as the possibility of growing dynamo modes from the velocity field
of a saturated nonlinear dynamo \citep{CT09}.
Even the $\Pm$ dependence of the dissipation ratio has already been
studied \citep{PS10,PSF13}.
Their results show the inverse dissipation ratio in semi-logarithmic
form, so the scaling for large $\Pm$ cannot be accurately assessed,
but their results are consistent with a constant dissipation ratio
for small $\Pm$ and show a sub-linear increase at large $\Pm$.

\begin{figure}[t!]\begin{center}
\includegraphics[width=\columnwidth]{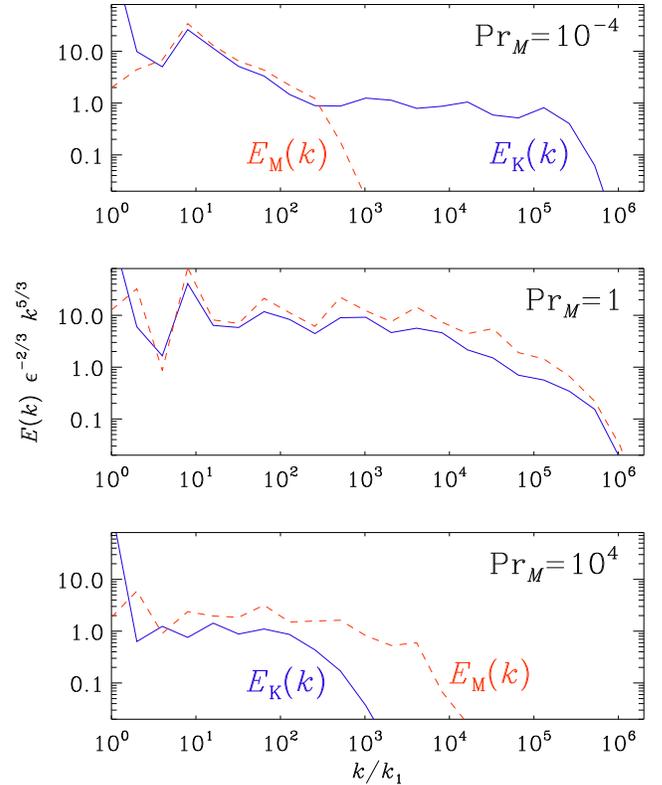}
\end{center}\caption[]{
Compensated time-averaged kinetic and magnetic energy spectra for
shell models at three values of $\Pm$.
}\label{pcomp_eps}\end{figure}

To assess the scaling more quantitatively, we now repeat their
calculations using an independent method.
The time integration is performed using an Adams-Bashforth scheme
with an integrating factor to treat the diffusion term \citep{BEO97}.
We use $N=30$ shells for $\Rey=u_0/\nu k_0$ of up to $10^9$ and
$\Pm$ in the range from $10^{-6}$ to $10^6$.
Forcing is applied by setting $F_1$ in \Eq{dundt} to a complex random number
at each time step, so this forcing is $\delta$-correlated, just like
in the DNS.
Compensated time-averaged spectra are shown in \Fig{pcomp_eps}
for three values of $\Pm$.
For $\Pm=1$, the magnetic and kinetic energy spectra are similar, while for
large (small) values of $\Pm$, the kinetic (magnetic) energy spectrum
is prematurely truncated, as is also the case in the DNS of \cite{B09}
for small $\Pm$.

\begin{figure}[t!]\begin{center}
\includegraphics[width=\columnwidth]{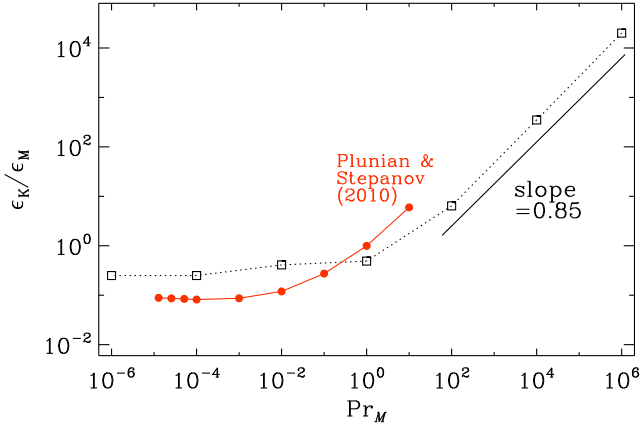}
\end{center}\caption[]{
$\Pm$ dependence of the dissipation ratio for the present shell models
(open squares), compared with the shell model results of \cite{PS10},
overplotted in red (filled circles).
}\label{shell_model_MHD_pepsKM}\end{figure}

As mentioned above, the $\Pm$ dependence of the dissipation ratio has
already been calculated by \cite{PS10}, and our present results
agree at least qualitatively with theirs.
In \Fig{shell_model_MHD_pepsKM} we show the $\Pm$ dependence of the
dissipation ratio $\epsK/\epsM$, where
\EQ
\epsK=2\nu\sum_{n=1}^N k_n^2|u_n|^2,\quad
\epsM=2\eta\sum_{n=1}^N k_n^2|b_n|^2.
\EN
The present shell models predict the dissipation ratio to be independent
of $\Pm$ for $\Pm<1$, which is in conflict with the DNS of \cite{B09}
where this trend was found to continue down to $\Pm=10^{-3}$.
On the other hand, the results of \cite{PS10}, which are overplotted in
\Fig{shell_model_MHD_pepsKM}, suggest a constant dissipation
ratio for $\Pm\la0.01$ only, which is already outside the plot range
of the present DNS shown in \Fig{pepsKM}, but still in conflict with
the DNS of \cite{B09} down to $\Pm=10^{-3}$.

For small values of $\Pm$, the present shell models show that the kinetic
energy cascade proceeds essentially independently of the magnetic field,
just like in ordinary hydrodynamic turbulence.
As explained in the introduction, this is also what one might have
naively expected, and it is perhaps even more surprising that this is
not borne out by the DNS.
On the other hand, for large values of $\Pm$, there is actually a
fairly strong $\Pm$ dependence, which is a direct consequence of $\epsM$
decreasing for large $\Rm$ rather than a consequence of $\epsK$ increasing.
Similar results have also been found for a one-dimensional passive scalar
model, which will be discussed next and compared with a one-dimensional
MHD model, which is an active scalar.

\section{Dissipation ratio in driven one-dimensional models}

The purpose of this section is to explore the possible behaviors
of simple models in which the spatial extent is fully resolved,
at least in one dimension.
Hydrodynamics in one dimension usually involves shocks,
such as the Burgers shock.
The pressureless idealization of the hydrodynamic equations is known as
the Burgers equation, and solutions can be found in closed form using
the Cole-Hopf transformation.
Before turning to the magnetic case, we should note that the evolution
of a passive scalar field in the presence of a Burgers shock was
already considered by \cite{OD10}, who found similar scaling to ours
in the limit of a large Schmidt number as $\Sc=\nu/\kappa$, where
$\kappa$ is the passive scalar diffusivity.

\subsection{Passive scalar model for a Burgers shock}

The passive scalar equation is a simple advection-diffusion equation
given by
\EQ
{\DD c\over\DD t}=\kappa\nabla^2c,
\EN
where $c$ is the passive scalar concentration and a relevant passive
scalar dissipation is defined as $\epsC=\bra{\kappa(\nabla c)^2}$.
For $\Sc\gg1$, \cite{OD10} found $\epsK/\epsC\propto\Sc^{1/2}$,
which is in remarkable agreement with the earlier findings for hydromagnetic
turbulence \citep{B09}.

Specifically, the equations considered by \cite{OD10} are
\EQA
\partial u/\partial t&=&-uu'+\tilde{\nu}u'',
\label{Burgers}\\
\partial c/\partial t&=&-uc'+\kappa c'',
\label{BurgersC}
\ENA
where primes denotes differentiation with respect to $x$.
The solution to \Eq{Burgers} decouples and possesses a shock.
In a frame of reference moving with the shock, the solution is
stationary and given by
\EQ
u(x)=-u_0\tanh x/w,
\EN
where $u_0$ is the velocity jump and $w_u=2\tilde{\nu}/u_0$ is the width
of the shock with $\tilde{\nu}=4\nu/3$ as a rescaled viscosity.
These equations can be obtained from the hydrodynamic version
(i.e., $\BB=\bm{0}$) of \Eq{Du} after setting $\cs=0$,
so the density gradient does not enter, and therefore we
can ignore \Eq{DH} and set $\rho=1$.
The $4/3$ factor in the expression for $\tilde{\nu}$ comes from the
fact that, owing to compressibility, the viscous acceleration term
includes a $\onethird\nab\nab\cdot\uu$ term in addition to the usual
$\nu\nabla^2\uu$ term; see \Eq{nab2S} for a corresponding reformulation
of the dissipation terms.
The viscous dissipation $\epsK=\tilde\nu\int (u')^2\,dx/L$,
using $\partial u/\partial x\propto1/\cosh^2(x/w)$, is then
\EQ
\epsK=\tilde\nu{w\over L}\int{dx/w\over\cosh^4(x/w)}
={4\over3}{\tilde{\nu}u_0^2\over wL}
={2\over3}{u_0^3\over L},
\EN
but here the $4/3$ factor comes from the fact that $\int\dd\xi/\cosh^4\xi=4/3$.
It is important to note that $\epsK$ is constant and independent of $\nu$.

On physical grounds, the passive scalar concentration is positive
definite.
Mathematically, however, \Eq{BurgersC} is invariant under the addition
of a constant.
We can therefore formulate the same boundary conditions for $c$
as for $u$, i.e., $c=\pm c_0$ and $u=\pm u_0$ for $x\to\mp\infty$,
which is truncated here at finite boundary positions $x=x_\mp$ that are
chosen to be sufficiently far away from the shock, i.e., $|x_\pm|\gg w$.

\begin{figure}[t!]\begin{center}
\includegraphics[width=\columnwidth]{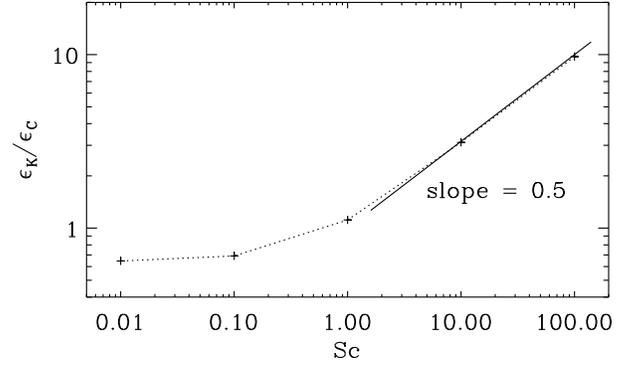}
\end{center}\caption[]{
Schmidt number dependence for the passive scalar case.
}\label{presults}\end{figure}

\begin{figure}[t!]\begin{center}
\includegraphics[width=\columnwidth]{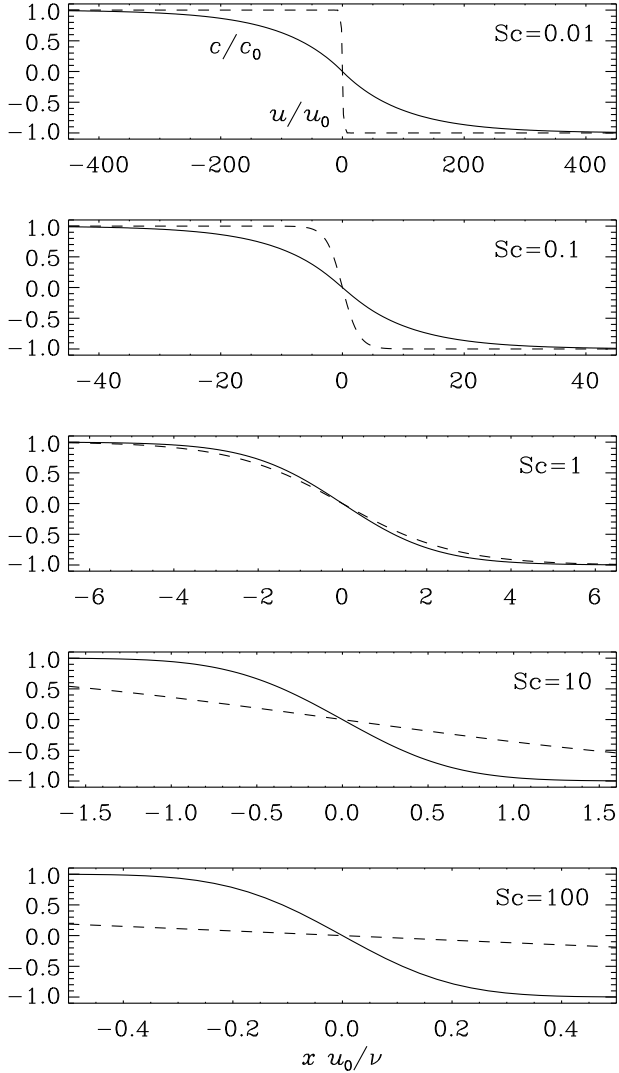}
\end{center}\caption[]{
Profiles of $c(x)$ (solid) and $u(x)$ (dashed) for different values of $\Sc$.
Note that the $x$ range decreases with increasing values of $\Sc$ so as
to have a similar coverage of the $c(x)$ profiles in all cases.
Note that we have scaled $x$ by $u_0/\nu$, so the hydrodynamic kink
always has the same width.
}\label{pprofC_comp}\end{figure}

However, for $\Sc\ll1$, one finds $\epsK/\epsC\approx\const$.
The dependence of $\epsK/\epsC$ as a function of $\Sc$ is shown in
\Fig{presults}, where we present the results from numerical integration.
There are clearly two different scalings for $\Sc\ll1$ and $\Sc\gg1$.
The profiles of $c(x)$ are shown in \Fig{pprofC_comp} for different
values of $\Sc$ and are compared with the profile of $u(x)$.
Not surprisingly, for small values of $\Sc$, the width of the kink
of $c(x)$ becomes wider.

\subsection{MHD model for Alfv\'en kinks}

An extension of the Burgers equation to MHD was already studied by
\cite{Tho68} and \cite{Pou93}, but unlike their cases which try to
model the effects of three-dimensional dynamos, here we employ just
a one-dimensional reduction of the three-dimensional equations to one
dimension, which results in equations equivalent to those of \cite{BNP14}.
This essentially implies a different sign in front of what corresponds to the
stretching term in MHD, i.e., the $\BB\cdot\nab\BB$ and $\BB\cdot\nab\uu$
non-linearities in the momentum and induction equations, respectively.
Unlike the case of a passive scalar, the magnetic field is an active
(vector) field which therefore back-reacts on the flow via the
Lorentz force, which in this case is just the magnetic pressure.
As before, the gas pressure is neglected ($\cs=0$), so
the governing equations therefore reduce to
\EQA
\partial u/\partial t&=&-uu'-bb'+\tilde{\nu}u'',\\
\partial b/\partial t&=&-ub'-bu'+\eta b'',
\ENA
These equations obey similar conservation equations as the
full MHD equation, except that here the energy input comes from
non-vanishing inflow at $x\to-\infty$ and is equal to $u_0^3/3L$.
Note, however, that there is no net Poynting flux, because $ub^2=0$
on both boundaries.

\begin{figure}[t!]\begin{center}
\includegraphics[width=\columnwidth]{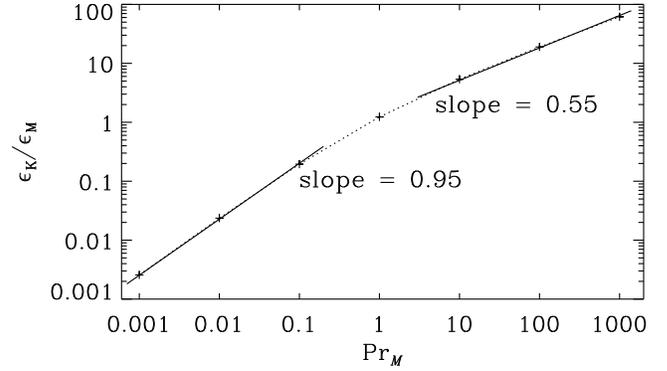}
\end{center}\caption[]{
Magnetic Prandtl number dependence in the MHD model.
}\label{pMresults}\end{figure}

\begin{figure}[t!]\begin{center}
\includegraphics[width=\columnwidth]{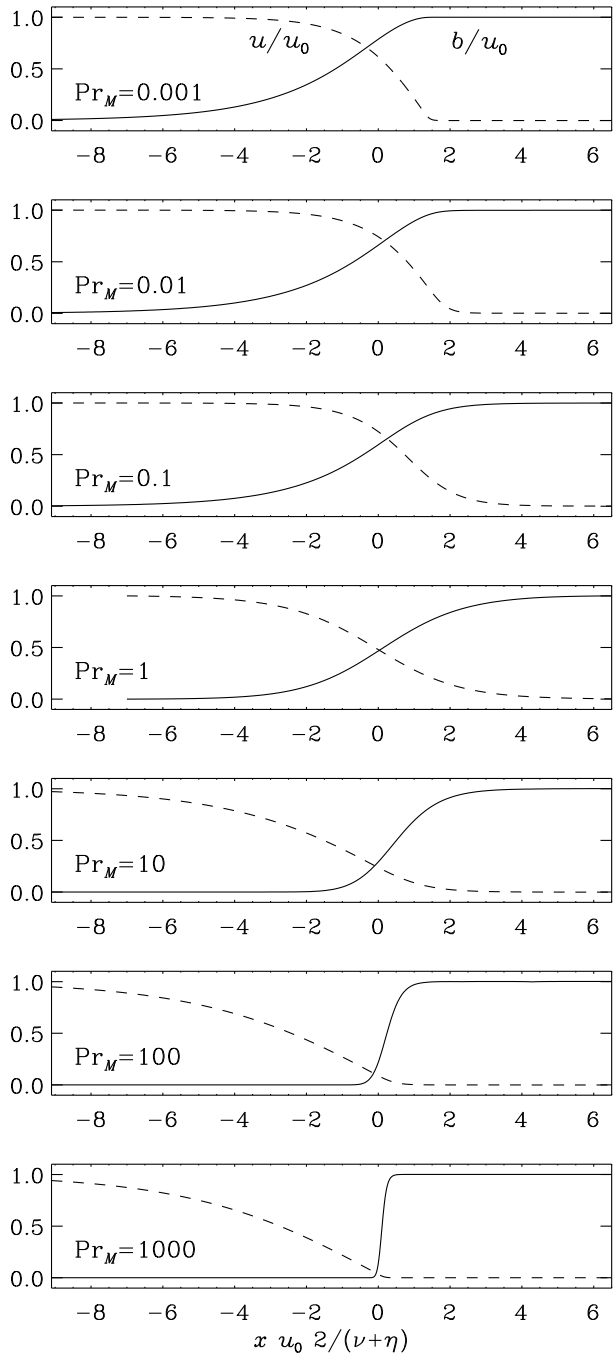}
\end{center}\caption[]{
Profiles of $b(x)$ (solid) and $u(x)$ (dashed) for different values of $\Pm$.
Note that the $x$ range is the same for all panels and that
we have normalized $x$ by $2u_0/(\nu+\eta)$.
}\label{pprofM_comp}\end{figure}

The magnetic cases are quite different from the passive scalar case
in that the magnetic field exerts a magnetic pressure.
One can therefore produce a stationary state where the ram pressure
of the flow from the left ($x\to-\infty$) can be balanced by
the magnetic pressure of a magnetic kink when $b\to u_0$ for $x\to+\infty$
and $b\to0$ for $x\to-\infty$.
Indeed, the stationary state must obey the following system of
two ordinary differential equations
\EQA
\label{ODEu}\partial u/\partial x&=&(u^2+b^2-u_0^2)/2\tilde\nu,\\
\label{ODEb}\partial b/\partial x&=&ub/\eta.
\ENA
In practice, however it was more straightforward to obtain solutions
using direct time integration in $x_-\leq x\leq x_+$ rather than
solving a two-point boundary value problem.
The resulting scaling in \Fig{pMresults} confirms \Eq{eps_scaling}
with $q\approx0.55$ for $\Pm>1$ and $q\approx0.95$ for $\Pm<1$.

Let us now discuss the profiles of $b(x)$ and $u(x)$ in the magnetic case,
shown in \Fig{pprofM_comp}.
Here we find scalings that are broadly similar to those for turbulent
large-scale dynamos as well as small-scale dynamos for $\Pm<1$,
namely a slope between $0.6$ and $0.7$.
For $\Pm=1$, the profiles of $b(x)$ and $u(x)$ are similar and resemble
the $\tanh x/w$ profile of $u$ in the passive scalar case.
However, for both $\Pm\ll1$ and $\gg1$, the profiles of $b(x)$ and $u(x)$
become asymmetric, which is also the reason why we chose
to integrate in a domain where $-x_->x_+$.
For small values of $\Pm$, i.e., when $\eta\gg\nu$, the magnetic field
begins to ramp up slowly and quite far away from $x=0$.
This leads to a corresponding decline of $u(x)$.
On the other hand, for large values of $\Pm$, the value of $\nu$ ($\gg\eta$)
is so large that a certain imbalance of $u^2+b^2-u_0^2$ in \Eq{ODEu}
implies only a small slope in $u(x)$, so $|u'|$ must be small.

The crucial point for the magnetic case is that the widths of the
magnetic and velocity kinks are never very different from each other.
Therefore, as a {\em zeroth} approximation, we can say that
$\bra{2\SSSS^2}/\bra{\uu^2}$ is approximately as large as
$\bra{\mu_0\JJ^2}/\bra{\BB^2}$ or, in our one-dimensional case,
$\bra{(u')^2}/\bra{u^2}$ is approximately as large as $\bra{(b')^2}/\bra{b^2}$.
Given that in all cases $\EK\approx\EM$, this would imply that
$\epsK/\epsM\propto\nu/\eta\equiv\Pm$, i.e., we would expect
linear scaling with $\Pm$.
In this case, as we have emphasized before, the usual phenomenology of
hydrodynamic turbulence, in which a decrease of $\nu$ implies a
corresponding increase of dissipation, is not obeyed.

\section{Effect of rotation}

The conversion of kinetic into magnetic energy is of obvious
astrophysical significance.
In stars with outer convection zones, a certain fraction
of the kinetic energy flux is converted into magnetic energy
and is observable as X-ray flux \citep[for example,][]{Vil84}.
This leads to a scaling law that has been verified over
many orders of magnitude \citep{CHR09}.
As we have seen, this scaling law must be affected by $\Pm$,
although the value of $\Pm$ is approximately the same
for all late-type stars, so this cannot easily be observationally checked.
However, what has not been checked is whether the conversion
also depends on the rotation rate.

\begin{figure}[t!]\begin{center}
\includegraphics[width=\columnwidth]{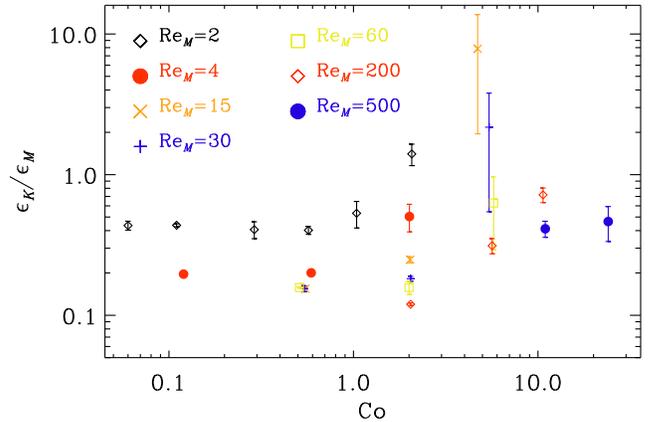}
\end{center}\caption[]{
Energy dissipation ratio as a function of the Coriolis number
for helically forced turbulence.
}\label{pepsKM_Co_dep}\end{figure}

In the work discussed in \Sec{TurbulentDynamos}, there was no explicit rotation.
Note, however, that \cite{PS10} did already study the effect of rotation
in their shell model calculations.
To check whether rotation influences our results,
we have performed a series of simulations with $\Co\neq0$
using $\Pm=1$, and have varied $\Rm$ ($=\Rey$) between 4 and 400,
and $\Co$ between 0.2 and 20.
The parameters of our runs are listed in \Tab{TsummaryR} and the result
is shown in \Fig{pepsKM_Co_dep}, where we plot $\epsK/\epsM$ as a function
of $\Co$.
The values of $\Rm$ are indicated by different symbols.

\begin{table*}\caption{
Summary of runs with $\Co\neq0$ and $\Pm=0.1$.
}\vspace{12pt}\centerline{\begin{tabular}{cccrrrcccccrrr}
\hline
\hline
Run & $\nu k_1/\cs$ & $\eta k_1/\cs$ & $\Rey$ & $\Rm$ & $\Co$ &
$\urms/\cs$ & $\brms/\cs$ & $\epsK/\epsT$ & $\epsM/\epsT$ & $C_\epsilon$ &
$k_\nu/k_1$ & $k_\eta/k_1$ & res.~~ \\
\hline
%
RA1  &$1\times10^{-3}$&$1\times10^{-2}$&  30&   3&  0.1& 0.090& 0.066&  0.30&  0.70&  1.71&  17&   4& $  16^3$ \\
RA2  &$1\times10^{-3}$&$1\times10^{-2}$&  30&   3&  0.1& 0.089& 0.064&  0.30&  0.70&  1.68&  17&   4& $  16^3$ \\
RA3  &$1\times10^{-3}$&$1\times10^{-2}$&  29&   3&  0.3& 0.088& 0.065&  0.29&  0.71&  1.66&  17&   4& $  16^3$ \\
RA4  &$1\times10^{-3}$&$1\times10^{-2}$&  29&   3&  0.6& 0.088& 0.065&  0.29&  0.71&  1.63&  17&   4& $  16^3$ \\
RA5  &$1\times10^{-3}$&$1\times10^{-2}$&  32&   3&  1.0& 0.096& 0.063&  0.35&  0.65&  1.31&  18&   4& $  16^3$ \\
RA6  &$1\times10^{-3}$&$1\times10^{-2}$&  40&   4&  2.1& 0.121& 0.053&  0.58&  0.42&  0.77&  21&   3& $  16^3$ \\
\hline
%
RB1  &$5\times10^{-4}$&$5\times10^{-3}$&  56&   6&  0.1& 0.084& 0.092&  0.16&  0.84&  2.10&  25&   7& $  32^3$ \\
RB2  &$5\times10^{-4}$&$5\times10^{-3}$&  57&   6&  0.6& 0.085& 0.094&  0.17&  0.83&  2.03&  25&   7& $  32^3$ \\
RB3  &$5\times10^{-4}$&$5\times10^{-3}$&  83&   8&  2.0& 0.124& 0.073&  0.34&  0.67&  0.62&  30&   6& $  32^3$ \\
\hline
%
RC1  &$2\times10^{-4}$&$2\times10^{-3}$& 150&  15&  0.6& 0.090& 0.117&  0.13&  0.87&  1.74&  48&  14& $  64^3$ \\
RC2  &$2\times10^{-4}$&$2\times10^{-3}$& 205&  21&  2.0& 0.123& 0.100&  0.20&  0.80&  0.66&  52&  13& $  64^3$ \\
RC3  &$2\times10^{-4}$&$2\times10^{-3}$& 353&  35&  4.7& 0.212& 0.019&  0.89&  0.11&  0.07&  65&   7& $  64^3$ \\
\hline
%
RD1  &$1\times10^{-4}$&$1\times10^{-3}$& 310&  31&  0.5& 0.093& 0.119&  0.13&  0.87&  1.56&  80&  23& $ 128^3$ \\
RD2  &$1\times10^{-4}$&$1\times10^{-3}$& 410&  41&  2.0& 0.123& 0.127&  0.15&  0.85&  0.69&  83&  23& $ 128^3$ \\
RD3  &$1\times10^{-4}$&$1\times10^{-3}$& 613&  61&  5.4& 0.184& 0.037&  0.69&  0.31&  0.12& 105&  16& $ 128^3$ \\
\hline
%
RE1  &$5\times10^{-5}$&$5\times10^{-4}$& 647&  65&  0.5& 0.097& 0.123&  0.14&  0.86&  1.44& 137&  39& $ 256^3$ \\
RE2  &$5\times10^{-5}$&$5\times10^{-4}$& 833&  83&  2.0& 0.125& 0.134&  0.14&  0.86&  0.69& 138&  39& $ 256^3$ \\
RE3  &$5\times10^{-5}$&$5\times10^{-4}$&1160& 116&  5.8& 0.174& 0.099&  0.39&  0.61&  0.17& 160&  32& $ 256^3$ \\
\hline
%
RF1  &$2\times10^{-5}$&$2\times10^{-4}$&2033& 203&  2.0& 0.122& 0.116&  0.11&  0.89&  0.59& 243&  74& $ 256^3$ \\
RF2  &$2\times10^{-5}$&$2\times10^{-4}$&2950& 295&  5.6& 0.177& 0.106&  0.24&  0.76&  0.14& 272&  65& $ 256^3$ \\
RF3  &$2\times10^{-5}$&$2\times10^{-4}$&3917& 392& 10.6& 0.235& 0.084&  0.42&  0.58&  0.06& 318&  62& $ 256^3$ \\
\hline
%
RG1  &$1\times10^{-5}$&$1\times10^{-4}$&7600& 760& 10.9& 0.228& 0.091&  0.29&  0.71&  0.07& 501& 111& $ 512^3$ \\
RG2  &$1\times10^{-5}$&$1\times10^{-4}$&6933& 693& 24.0& 0.208& 0.089&  0.32&  0.68&  0.09& 496& 107& $ 512^3$ \\
\hline
\hline\label{TsummaryR}\end{tabular}}
\end{table*}

We see that for a given value of $\Rm$, there is a certain
value of $\Co=\Co^*$ below which $\epsK/\epsM$ is roughly unaffected
by rotation.
As the value of $\Rm$ is increased, $\Co^*$ also increases,
thereby extending the range over which $\epsK/\epsM$ remains
roughly independent of $\Co$.
In astrophysical applications, $\Rey$ is usually large enough so that we
should not expect to see any rotational dependence of $\epsK/\epsM$.
This explains why the scaling result of \cite{CHR09}
follows the expected scaling of $\epsK+\epsM\approx\urms^3/L$ with some
length scale $L$ over a huge range.

\section{Conclusions}

In the present work, he have extended earlier findings of a $\Pm$ dependence
of the kinetic-to-magnetic energy dissipation ratio, $\epsK/\epsM$,
to the regime of small-scale and large-scale dynamos for $\Pm>1$
and at higher resolution than what was previously possible \citep{B11an}.
In most cases, our results confirm earlier results that for large-scale dynamos,
the ratio $\epsK/\epsM$ is proportionate to $\Pm^{0.6}$.
Furthermore, we have shown that a similar scaling with $\Pm$ can be
obtained for a simple one-dimensional Alfv\'en kink, where ram pressure
locally balances magnetic pressure.
Interestingly, in these cases kinetic energy dissipation is accomplished
mainly by the irrotational part of the flow rather than the solenoidal part
as in the turbulence simulations presented here.
We note in this connection that the kinetic energy dissipation,
which is proportional to
$\bra{2\SSSS^2}=\bra{(\nab\times\uu)^2}+\bra{{4\over3}(\nab\cdot\uu)^2}$,
has similar contributions from vortical and irrotational parts.

We have also shown that for fixed values of $\Pm$, the ratio
$\epsK/\epsM$ is not strongly dependent on the presence of rotation,
provided the magnetic Reynolds number is not too close to the marginal value
for the onset of dynamo action.
In the simulations with $\Co\neq0$ presented here, the runs were often
not very long and therefore the error bars large, but the number of
similar results support our conclusions that $\epsK/\epsM$ is roughly
independent of $\Co$.

For many astrophysical systems, the microscopic energy dissipation
mechanism is not of Spitzer type, as assumed here.
It is not obvious how this would affect our results.
Nevertheless, it is clear that conclusions based on the
kinetic-to-magnetic energy ratio itself do not have much bearing
on the energy {\em dissipation} ratio.
This became clear some time ago in connection with local accretion
disk simulations driven by the magneto-rotational instability, where
magnetic energy strongly dominates over kinetic.
However, as it turned out, most energy is dissipated viscously rather than
resistively \citep{BNST95}.

Unfortunately, the question of energy dissipation is not routinely examined
in astrophysical fluid dynamics, nor is it always easy to determine energy
dissipation rates,
because many astrophysical fluid codes ignore explicit dissipation and
rely entirely on numerical prescriptions needed to dissipate energy when
and where needed.
Our present work highlights once again that this can be a questionable
procedure, because it means that even non-dissipative aspects, such as
the strength of the dynamo which is characterized by
$\bra{\uu\cdot(\JJ\times\BB)}$, are then ill-determined.
The reason why this has not been noted earlier is that most previous
work assumed $\Pm$ to be of the order of unity.
An exception is the work of \cite{B09}, where dynamo simulations for
values of $\Pm$ as small as $10^{-3}$ were considered.
One reason why such extreme values of $\Pm$ have been possible is the fact that
at very small values
of $\Pm$, most of the energy is dissipated resistivity, and there
is not much kinetic energy left at the end of the turbulent kinetic
energy cascade.
As a consequence, it is then possible to decrease the value of $\nu$
further and still dissipate the remaining kinetic energy,
which implies that the nominal value of the fluid Reynolds
number can become much larger than what is usually possible when there is
no additional resistive dissipation.
However, one may wonder how a large-scale dynamo can depend on $\Pm$.
We expect that this is only possible if most of the energy transfer
comes ultimately from small scales.

It is also noteworthy that there is now some evidence for the non-universal
behavior of the scaling of the kinetic-to-magnetic energy dissipation ratio
with $\Pm$.
Although some of the earlier results with slightly different exponents
could be explained by inaccuracies and other physical effects, there are
now examples such as one-dimensional simulations and the passive scalar
analogy that display different
exponents which cannot easily be explained through artifacts.
Also, the result that for large enough magnetic Reynolds numbers
the dissipation ratio scales differently in the presence of helicity
($q\approx0.7$) than without ($q\approx1/3$) is surprising.
It would therefore be interesting to revisit the viscous-to-magnetic
dissipation ratios over a broader range of circumstances.

\acknowledgments
I thank Koji Ohkitani for useful discussions and Rodion Stepanov for
comments on the paper and for providing the dissipation rates found in
his paper with Franck Plunian, which are now overplotted in Figure~6 of
the present paper.
I also thank Prasad Perlekar for referring me to the data of their
paper of 2011, which are now overplotted in Figure~3 of the present paper.
This work was supported in part by
the European Research Council under the AstroDyn Research Project No.\ 227952,
and the Swedish Research Council grants No.\ 621-2011-5076 and 2012-5797,
as well as the Research Council of Norway under the FRINATEK grant 231444.
We acknowledge the allocation of computing resources provided by the
Swedish National Allocations Committee at the Center for
Parallel Computers at the Royal Institute of Technology in
Stockholm and the National Supercomputer Centers in Link\"oping,
the High Performance Computing Center North in Ume\aa,
and the Nordic High Performance Computing Center in Reykjavik.


\end{document}